\documentclass[epjST]{svjour}
\usepackage{graphicx}
\begin{document}
\title{Superfluid density and compressibility at the superfluid-Mott glass 
      transition}

\author{Cameron Lerch \and Thomas Vojta\thanks{\email{vojtat@mst.edu}}  }
\institute{Department of Physics, Missouri University of Science and Technology, Rolla, Missouri 65409, USA}
\abstract{
Systems of disordered interacting bosons with particle-hole symmetry
can undergo a quantum phase transition between the superfluid phase
and the Mott glass phase which is a gapless incompressible insulator.
We employ large-scale Monte Carlo simulations of a two-dimensional
site-diluted quantum rotor model to investigate the properties of
the superfluid density and the compressibility at this transition.
We find that both quantities feature power-law critical behavior
with exponents governed by generalized Josephson relations.
} 
\maketitle

\section{Introduction}
\label{sec:intro}

The physics of systems as diverse as cold atoms in disordered optical
lattices \cite{WPMZCD09,KSMBE13,DTGRMGIM14}, superconducting thin films
\cite{HavilandLiuGoldman89,HebardPaalanen90}, Josephson
junction arrays \cite{ZFEM92,ZEGM96}, helium absorbed in vycor
\cite{CHSTR83,Reppy84}, and doped quantum magnets in large magnetic fields
\cite{OosawaTanaka02,HZMR10,Yuetal12} can be described by models of
disordered and interacting bosons. These systems can undergo zero-temperature
quantum phase transitions between superfluid and insulating ground states.

Because of the disorder, the conventional bulk phases, viz., superfluid
and Mott insulator, are separated by a quantum Griffiths phase
\cite{Griffiths69,ThillHuse95,Vojta06,Vojta10} in which
superfluid ``puddles'' exist in an insulating matrix. Depending on the symmetries,
this Griffiths phase can either be a
Bose glass or a Mott glass. The Bose glass phase, a compressible gapless
insulator, occurs for generic disorder without particle-hole symmetry
\cite{FisherFisher88,FWGF89,PPST09}. If the disordered Hamiltonian
is particle-hole symmetric, the glassy phase between the superfluid and
the Mott insulator is the incompressible, gapless, and insulating
Mott glass (sometimes also called random-rod glass)
 \cite{GiamarchiLeDoussalOrignac01,WeichmanMukhopadhyay08}.

The zero-temperature quantum phase transition between
superfluid and Mott glass was recently investigated by large-scale
Monte Carlo simulations \cite{Vojtaetal16,PuschmannVojta17},
resolving contradictions between earlier predictions
\cite{ProkofevSvistunov04,IyerPekkerRefael12,SLRT14}. Employing finite-size
scaling of the order parameter as well as the correlation length and time,
the critical behavior was found to be of conventional power-law type,
with universal critical exponents. This agrees with the general classification
of phase transitions in disordered systems based on the effective
dimensionality of the defects \cite{Vojta06,Vojta10,VojtaSchmalian05}.

In the present paper, we focus on the behavior of two experimentally important quantities,
the superfluid density $\rho_s$ and the compressibility $\kappa$, at this quantum
phase transition. By means of Monte Carlo simulations, we establish that both
quantities feature power-law critical behavior. The corresponding critical
exponents fulfill generalized Josephson relations \cite{FWGF89}.
The paper is organized as follows: In Sec.\ \ref{sec:model}, we define
the model and the quantities under consideration.
The Monte-Carlo simulation results are presented in Sec.\ \ref{sec:results}.
We conclude in Sec.\ \ref{sec:conclusions}.

\section{Quantum rotor model, superfluid density, and compressibility}
\label{sec:model}

We consider a square-lattice site-diluted quantum rotor model defined by
the Hamiltonian
\begin{equation}
 H = \frac U 2 \sum_i \epsilon_i \hat n_i^2 -J\sum_{\langle ij \rangle} \epsilon_i \epsilon_j \cos(\hat \phi_{i}-\hat \phi_j)~.
\label{eq:Hamiltonian}
\end{equation}
Here, $\hat n_i$ denotes the number operator of lattice site $i$ and  $\hat \phi_i$ the canonically
conjugate phase operator. $U$ and $J$ are the charging energy and the Josephson coupling, respectively,
and $\langle ij \rangle$ denotes pairs of nearest neighbor sites.
The site dilution is implemented via the independent quenched random variables
$\epsilon_i$ that can take the values 0 (vacancy) with probability $p$ and 1
(occupied site) with probability $1-p$.
If the filling, i.e., the average particle number $\langle n \rangle$, is an integer then the Hamiltonian is particle-hole
symmetric, which is the case we focus on in the following.
The qualitative behavior of the rotor model (\ref{eq:Hamiltonian}) is easily understood:
If the charging energy dominates, $U\gg J$, the ground state is Mott-insulating.
For $U\ll J$ and dilutions below the percolation threshold, the ground state is superfluid.

We now map the quantum rotor Hamiltonian (\ref{eq:Hamiltonian})  onto a classical (2+1)-dimensional XY model
\cite{WSGY94} with columnar defects,
\begin{equation}
 H_{\rm cl} = -J_{s}\sum_{\langle i,j\rangle, t}
\epsilon_i\epsilon_j\mathbf{S}_{i,t}\cdot\mathbf{S}_{j,t}
-J_\tau \sum_{i,t}\epsilon_i\mathbf{S}_{i,t}\cdot\mathbf{S}_{i,t+1}~.
\label{eq:Hcl}
\end{equation}
Here $i$ denotes a position in two-dimensional real space, and $t$ is the ``imaginary time'' coordinate.
$\mathbf{S}_{i,t}$ is an O(2) unit vector.
The values of the classical interactions $J_{s}/T$ and $J_\tau/T$ depend on the parameters
of the quantum Hamiltonian (\ref{eq:Hamiltonian}). The classical temperature $T$ differs from the real physical temperature $T_Q$
which is zero at the quantum phase transition.
As the critical behavior is expected to be universal, we fix $J_{s}=J_\tau=1$
and tune the transitions by varying the classical temperature $T$.

Under the quantum-to-classical mapping, the compressibility $\kappa=\partial \langle n \rangle /\partial \mu$
of the quantum rotor Hamiltonian (\ref{eq:Hamiltonian}) maps
onto the spinwave stiffness in the imaginary time direction
\begin{equation}
\rho_{cl,\tau} = L_\tau^2 (\partial^2 f / \partial \Theta^2)_{\Theta=0}
\label{eq:stiff_time}
\end{equation}
of the XY model (\ref{eq:Hcl})  (up to constant factors). Here, $L_\tau$ is the system size in
imaginary time direction, and $f$ is the free energy density for twisted boundary conditions
(the XY spins $\mathbf{S}_{i,t}$ at $t=0$ make an angle of $\Theta$ with those at $t=L_\tau$).
Analogously,
the superfluid density $\rho_s$ of the quantum rotor model maps onto the spinwave stiffness
\begin{equation}
\rho_{cl,s} = L^2 (\partial^2 f / \partial \Theta^2)_{\Theta=0}
\label{eq:stiff_space}
\end{equation}
in the space direction of the XY model, with $L$ being the spatial linear system size. (The twisted boundary
conditions are now applied in $x$ or $y$ direction.)

\section{Monte Carlo simulations}
\label{sec:results}

We perform Monte Carlo simulations of the classical XY model (\ref{eq:Hcl}) combining
Wolff \cite{Wolff89} and Metropolis updates \cite{MRRT53}. Wolff cluster
updates greatly reduce critical slowing down while Metropolis single-spin updates
help equilibrate disconnected clusters that occur in a diluted lattice. We study a dilution
of $p=1/3$ because it was found to lead to small corrections
to scaling at the disordered critical point \cite{Vojtaetal16}. For comparison, we also
analyze the clean case, $p=0$.
System sizes range from $L=10$ to 84 in the space directions and
from $L_\tau=14$ to 348 in imaginary time direction. All data are averaged over 10,000
disorder configurations. The equilibration period is 200 Monte Carlo sweeps for each sample,
and the measurement period is 500 sweeps, with a measurement taken after each sweep.
(Performing short Monte Carlo runs for a large number of disorder configurations reduces
the overall statistical error \cite{BFMM98,BFMM98b,SknepnekVojtaVojta04,VojtaSknepnek06,ZWNHV15}).

We compute the temporal and spatial stiffnesses $\rho_{cl,\tau} $ and $\rho_{cl,s}$ by employing numerical estimators
that can be evaluated in conventional Monte Carlo runs without actually having to apply
twisted boundary conditions \cite{TeitelJayaprakash83}. To test our algorithms, we first analyze
the clean case, $p=0$, using systems with up to $224^3$ sites. As the clean system is isotropic,
$\rho_{cl,\tau} $ and $\rho_{cl,s}$ agree with each other. At the critical temperature,
$T_c=2.201844$ \cite{Vojtaetal16}, they decay as $L^{-1}$ with system size, as predicted by the
Josephson scaling relation $\rho_{cl} \sim L^{2-d_{cl}}$ \cite{Josephson66}. Here, $d_{cl}=d+1=3$
is the total dimensionality of the classical XY model (\ref{eq:Hcl}).

\begin{figure}
\includegraphics[width=6cm]{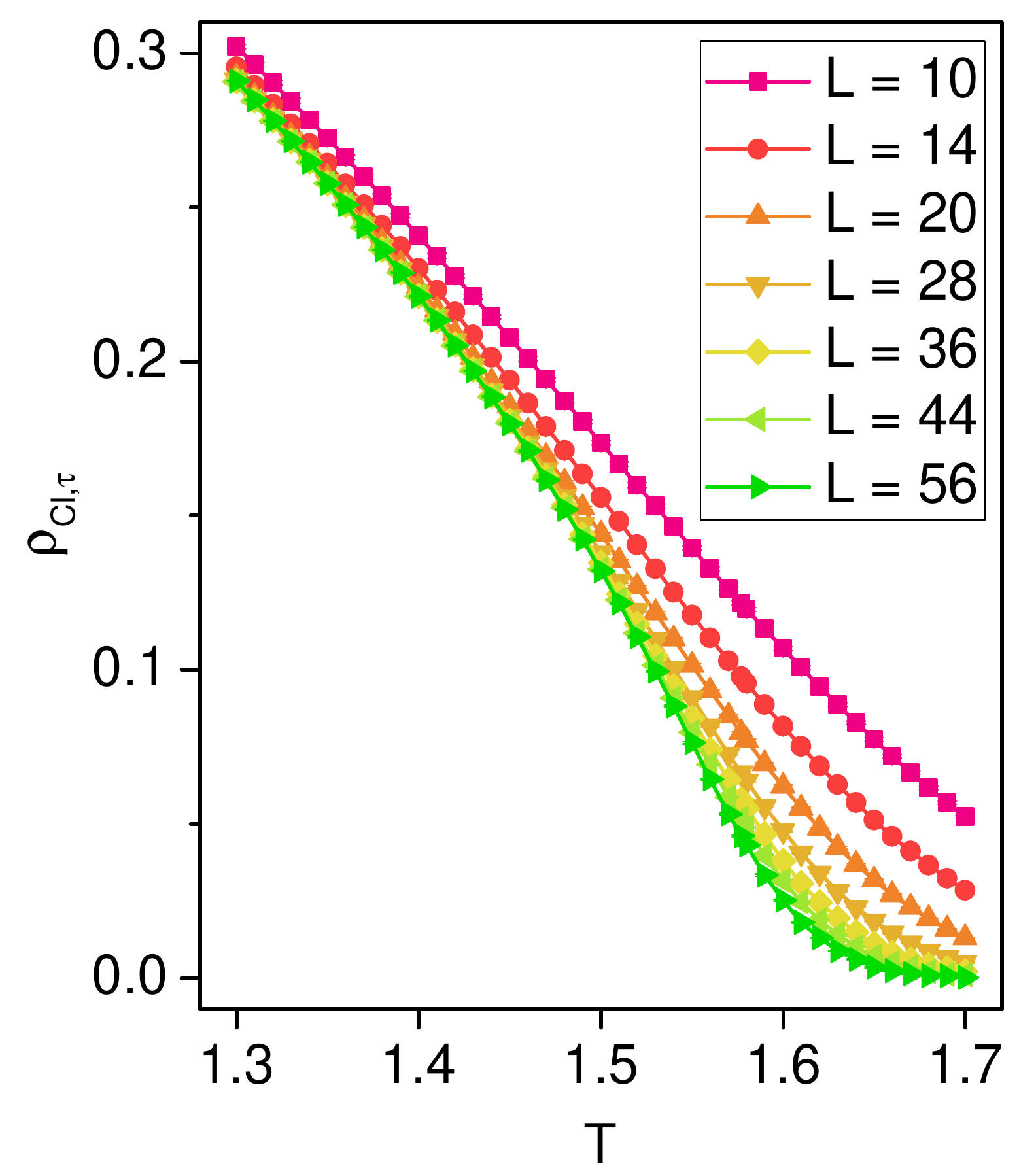}
\includegraphics[width=6cm]{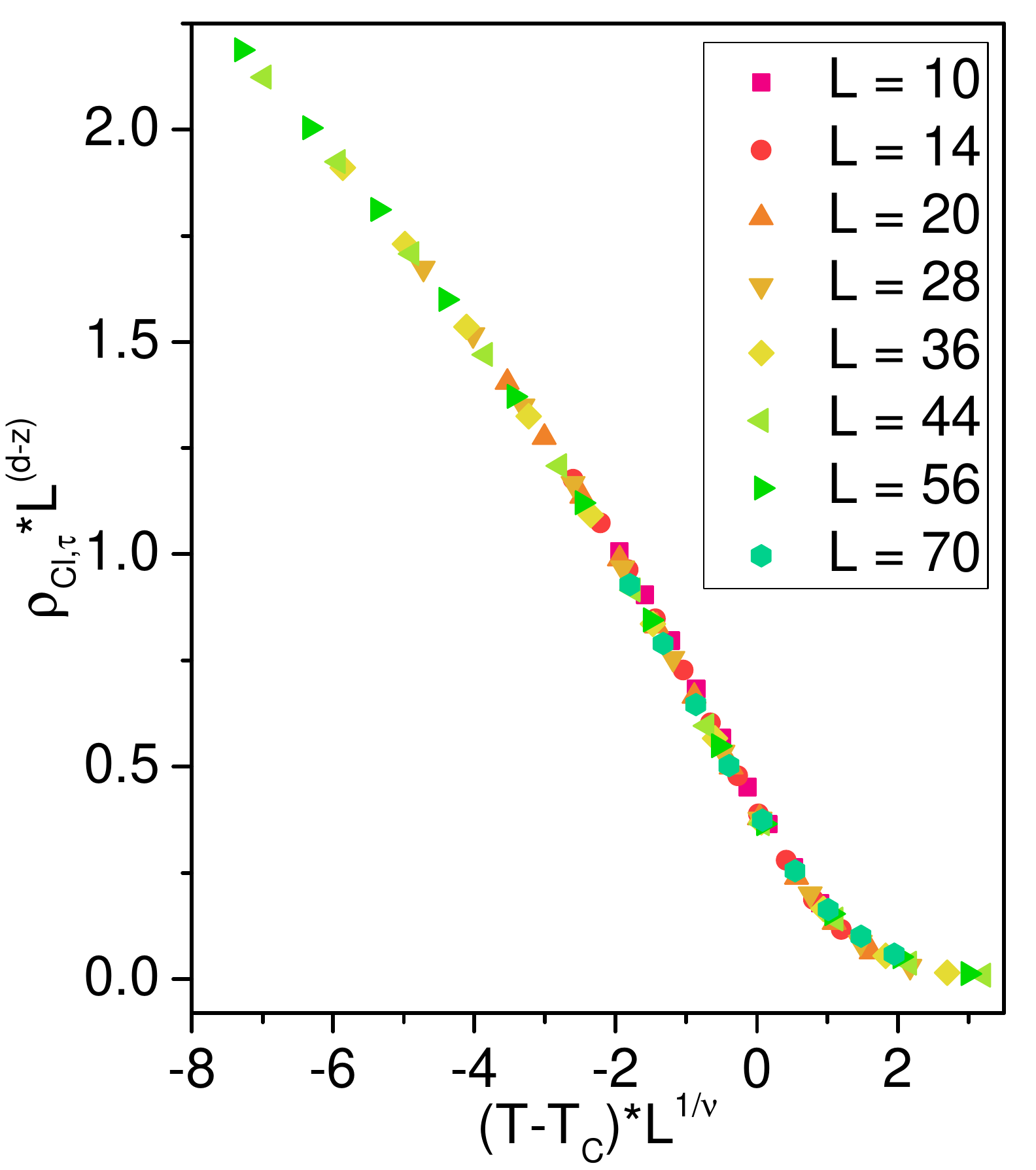}
\caption{Left: Stiffness $\rho_{cl,\tau}$ in imaginary time direction vs.\ classical temperature $T$
for various systems sizes. The dilution is $p=1/3$. The statistical errors of the data are  smaller
than the symbol size. Right: Scaling plot of $\rho_{cl,\tau}$ according to eq.\ (\ref{eq:stiff_scaling}).}
\label{fig:1}
\end{figure}

We now turn to the main results of this paper, viz., the behavior of $\rho_{cl,\tau} $ and $\rho_{cl,s}$
in the disordered case, $p=1/3$. As the disorder breaks the symmetry between space and imaginary time,
$L$ and $L_\tau$ are not equivalent, and we need to perform anisotropic finite-size scaling. The optimal
sample shapes for the present simulations as well as the value of the critical temperature, $T_c=1.577$,
are taken from Ref.\ \cite{Vojtaetal16}. The left panel of Fig.\ \ref{fig:1} presents an overview of
the temporal stiffness $\rho_{cl,\tau}$ for several system sizes.
As expected, $\rho_{cl,\tau}$ approaches a nonzero value for $T<T_c$ (superfluid phase) while it
decays towards zero for $T>T_c$ (Mott glass phase). The spatial stiffness $\rho_{cl,s}$ behaves in a similar
fashion.

\begin{figure}
\sidecaption
\includegraphics[width=8.5cm]{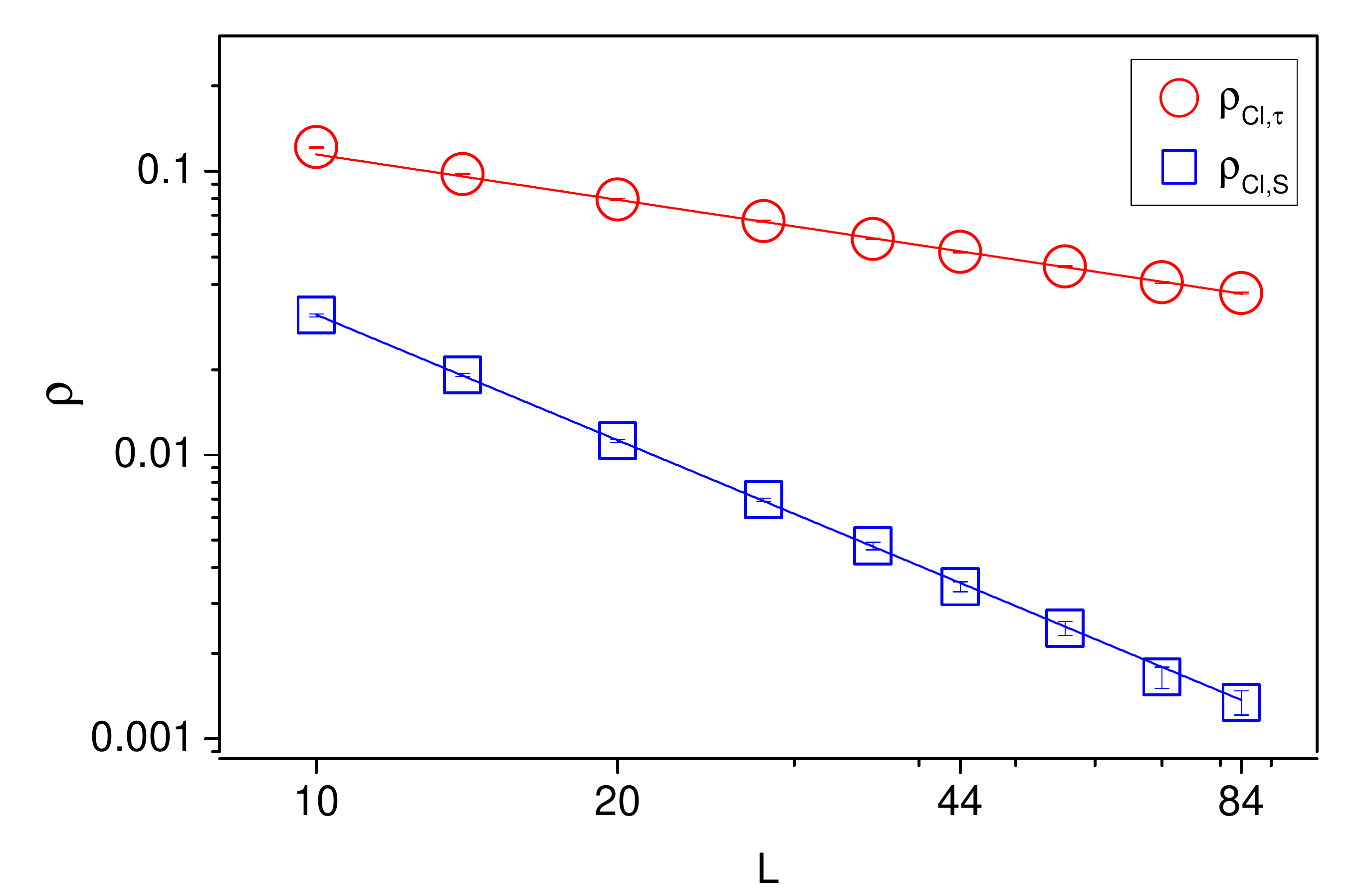}
\caption{Double logarithmic plot of the spinwave stiffnesses $\rho_{cl,\tau} $ and $\rho_{cl,s}$ at the critical temperature
$T_c=1.577$ as functions of the  spatial linear system size $L$. The solid lines are power-law fits to
$\rho_{cl,\tau} \sim L^{-y_\tau} $ and $\rho_{cl,s} \sim L^{-y_s}$.}
\label{fig:2}
\end{figure}

To determine the critical behavior of the stiffnesses $\rho_{cl,\tau} $ and $\rho_{cl,s}$ quantitatively,
we analyze their system-size dependence at the critical temperature $T_c=1.577$ in Fig.\ \ref{fig:2}.
The figure demonstrates that both stiffnesses feature power-law behavior,
$\rho_{cl,\tau} \sim L^{-y_\tau} $ and $\rho_{cl,s} \sim L^{-y_s}$ where
$y_\tau$ and $y_s$ are the scale dimensions of the stiffnesses.
Power-law fits yield the values
$y_\tau=0.53(5)$ and $y_s=1.47(6)$. The errors (indicated by the numbers in parentheses)
mostly stem from the uncertainty of $T_c$; the statistical errors are much smaller.
According to the generalized Josephson relations \cite{FWGF89}, $y_\tau$ and $y_s$
should fulfill the exponent equalities $y_\tau=d-z$ and $y_s=d+z-2$. Using $d=2$ and the estimate
$z=1.52(3)$ from Ref.\ \cite{Vojtaetal16}, we see that both relations are fulfilled within their
error bars.

A more complete analysis of the critical behavior is provided by testing the scaling
behavior of the stiffnesses. The scaling form of the
temporal stiffness reads
\begin{equation}
\rho_{cl,\tau}(r,L) = L^{-y_\tau} \, X_\tau(rL^{1/\nu})
\label{eq:stiff_scaling}
\end{equation}
where $\nu=1.16$ \cite{Vojtaetal16} is the correlation length exponent,
$r=T-T_c$ denotes the distance from criticality, and $X_\tau$ is a universal scaling function.
The right panel of Fig.\ \ref{fig:1} shows that our data fulfill the scaling
form with high accuracy. The spatial stiffness can be analyzed in a similar manner.

\section{Conclusions}
\label{sec:conclusions}

Couched in terms of the original disordered boson problem, i.e., the quantum rotor Hamiltonian
(\ref{eq:Hamiltonian}), our findings can be summarized as follows:
Close to the superfluid-Mott glass quantum phase transition, the compressibility
and the superfluid density  display power-law critical behavior,
$\kappa \sim L^{-y_\tau} $ and $\rho_{s} \sim L^{-y_s}$. The scale dimensions
$y_\tau=0.53(5)$ and $y_s=1.47(6)$ fulfill the generalized Josephson relations,
$y_\tau=d-z$ and $y_s=d+z-2$, within their error bars.

Combining these results with the critical exponents determined in Ref.\
\cite{Vojtaetal16}, we can write down the full scaling forms of $\kappa$ and $\rho_s$
as functions of the distance $r$ from criticality, temperature $T_Q$, and system size
$L$,
\begin{eqnarray}
\kappa(r,T_Q,L) &=& b^{-y_\tau} \, \kappa(rb^{1/\nu},T_Q b^z,L b^{-1})~, \\
\rho_s(r,T_Q,L) &=& b^{-y_s} \,    \rho_s(rb^{1/\nu},T_Q b^z,L b^{-1})~.
\end{eqnarray}
Here, $b$ is an arbitrary length scale factor.
(Recall that $T_Q$ is the physical temperature of the quantum system, not the
classical temperature $T$ appearing in the mapped classical XY model.)

Potential experimental realizations of Mott glass physics can be found, e.g., in
granular superconductors, ultracold atoms, and certain magnetic systems
\cite{RoscildeHaas07}. The last example is particularly
promising as the necessary particle-hole symmetry arises naturally
in the absence of a magnetic field.

\begin{acknowledgement}
This work was supported in part by the NSF under Grant Nos.\  DMR-1506152 and  PHY-1125915.
T.V. is grateful for the hospitality of the Kavli Institute
for Theoretical Physics, Santa Barbara, where part of the work was performed.
\end{acknowledgement}

\bibliographystyle{epj}
\bibliography{../00Bibtex/rareregions}

\end{document}